\documentstyle[preprint,aps,pra]{revtex}
\begin{document}
\draft
\title{Chaotic Signatures in the Spectrum of a Quantum Double Well}
\author{R. Berkovits$^{\text{(1)(2)}}$, Y. Ashkenazy$^{\text{(2)}}$, L. P.
Horwitz$^{%
\text{(2)(3)}}$and J. Levitan$^{\text{(2)(4)}}$}
\address{(1) The Jack and Pearl Resnick Institute of Advanced Technology,\\
(2) Department of Physics, Bar-Ilan University, Ramat-Gan 52900, Israel\\
(3) School of Physics, Raymond and Beverly Sackler Facuty of Exact Sciences\\
Tel-Aviv University, Ramat-Aviv, Israel.\\
(4) The Research Institute, The College of Judea and Samaria,\\
Kedumim and Ariel, P.O.B. 3 Ariel, 44837 Israel.}
\date{\today }
\maketitle

\begin{abstract}
The spectrum of a double well constructed of a square barrier embedded in an
infinite well is analyzed. Level statistics for levels slightly above the
barrier show signs of Wigner statistics usually associated with quantum
chaos. The correspondence with Wigner statistics improves when an ensemble
of systems with slightly different barrier heights is taken, possibly
reflecting an adiabatic time-dependent modulation of the barrier.
\end{abstract}


\narrowtext

One of the most celebrated definitions of classical chaos is the exponential
divergence in time of neighboring points in phase space. In discussing the
possible existence of quantum chaos, one would like to keep as close as
possible to the above definition. Implicitly, one needs to construct a
legitimate quantum mechanical paraphrase of the above definition. In fact,
the wave function of a particle may be thought of as representing an
ensemble of points in phase space while its spread as function of time is a
measure of the divergence of those points. We have recently considered the
time development of a wave packet initially located at the left hand side of
a square barrier embedded in an infinite well \cite{r1}. We found, in
addition to the highly complex behavior of the wave packet, that the spread
validity of the Ehrenfest approximation fails much more rapidly than for a
wave packet in free space.

Ballentine {\it et al} \cite{r2} have argued that a quantum state may behave
essentially classically, even when Ehrenfest's theorem does not apply. They
conclude that Ehrenfest's theorem is neither necessary nor sufficient to
identify the classical regime and that the classical limit of a quantum
state is not a single classical trajectory but rather an ensemble of
classical trajectories. However, Ehrenfest's theorem breaks down much sooner
for a chaotic ensemble than for a regular one.

Pattanyak and Schieve \cite{r3} have studied a one dimensional problem with
a Duffing potential without external perturbation. Their results reflect the
highly complex behavior of the quantum state even for a case in which the
system is {\it not} chaotic in the classical limit.

Another indicator commonly used in the study of quantum chaos is the
properties of the energy spectrum. It is well known \cite{richel,gut,metha}
that the energy level statistics of quantum systems show different
statistical characteristics corresponding to whether the system is chaotic
or regular in the classical limit \cite{r7}. For systems which are regular
in the classical limit the energy level statistics correspond to Poisson
statistics, while for systems which are chaotic in the classical limit the
statistics follow Gaussian ensemble statistics (known also as the Wigner
statistics) according to the system's symmetry. Systems with time reversal
symmetry follow the Gaussian orthogonal ensemble, and systems without time
reversal symmetry follow the Gaussian Unitary ensemble.

A useful statistical measure of the energy level statistics is the level
spacing distribution. The Poisson distribution is given by
\begin{equation}
P_P(s) = \exp (-s),  \label{e1}
\end{equation}
where $s$ is the level spacing in units of the averaged level spacing. For
the Gaussian orthogonal ensemble
\begin{equation}
P_W(s) = {\frac{{\pi s} }{{2}}} \exp \left( {\frac{{- \pi s^2} }{{4}}}
\right).  \label{e2}
\end{equation}

One can also characterize the level statistics by the spectral rigidity $%
\Delta _3(L)$. The spectral rigidity is defined as the local average of the
mean square deviation of the cumulative number of states $N(E)$ from the
best fitting straight line over a range of energy corresponding to $L$ mean
level spacings
\begin{equation}
\Delta _3(L)=\langle {\frac{{1}}{{L}}}\int_\alpha ^{\alpha +L}dE\big[ %
N(E)-AE-B\big]^2\rangle .  \label{stiff}
\end{equation}
For the Poisson statistics $\Delta _3(L)={\frac{{L}}{{15}}}$ while for the
Gaussian orthogonal ensemble $\Delta _3(L)={\frac{{1}}{{\pi ^2}}}(\ln
(L)-.0687)$. The spectral rigidity is often a stronger tool than the level
spacing distribution in the analysis of complex systems, since it takes into
account the correlations between levels on large energy scales while the
spacing distribution takes into account only nearest neighbor correlations.

In this paper we calculate the spectrum of a double well constructed of a
square barrier embedded in an infinite well. We have calculated the level
spacing distribution and the spectral rigidity for different regions of the
spectrum. For energies significantly below or above the barrier level we
find behavior consistent with simple square well level statistics. In the
neighborhood of the barrier edge we find statistics which are closer to the
Wigner statistics. When the spectral rigidity is averaged over several
different barrier heights (different from the original height by no more
than a level spacing), which has the effect of increasing the sample
population, the correspondence with the Wigner prediction is very good.
Thus, for energies in that region the spectrum exhibits a signature of
chaos. We emphasize however that the use of the energy level statistics as
an indication of chaos must be taken with a grain of salt since for the
system we are studying there is no classical limit \cite{r7}.

The energy levels were calculated for a square barrier of height $V=5$ and
width $2a=2$ ($\hbar \equiv 1,2m\equiv 1$) embedded in an infinite well of
width $2b=110$. The cumulative number of levels is presented in Fig. \ref
{fig.1}. As can be expected the low eigenvalues ($E<V$) appear as almost
degenerate pairs around $E_{2n}\sim E_{2n+1}\sim \hbar ^2(\pi n)^2/2m(b-a)^2$%
, which is the energy level for an infinite square well of width $b-a$. This
can be clearly seen in $N(E)$ which has a step like shape where each step is
equal to two. The spacings between the high eigenvalues ($E\gg V$)
correspond to the spacings expected from an infinite well of width $2b$,
i.e., the barrier has little influence on those levels. For energies just
above the barrier height $V<E<2V$ the behavior of $N(E)$ is more complex and
can not be described by any simple formula. Therefore, this region is a
natural candidate for a more sophisticated analysis of its statistical
properties.

After performing the standard unfolding procedure of the spectrum \cite
{richel,gut,metha}, the spectral rigidity $\Delta _3(L)$ for different
regions of the spectrum is calculated. The results are plotted in Fig. \ref
{fig.2}. For the high energy levels $\Delta _3(L)=1/12$ which is the
rigidity of equal spaced levels known as a ``picket fence''. This is to be
expected since the levels in this region have almost no local fluctuations,
and the global increase of $N(E)\sim \sqrt{E}$ is removed by the unfolding.
A similar situation is seen for the low-lying levels with a small difference
which stems from the fact that those levels are composed of sets of two
degenerate levels equally spaced, i.e., a double picket fence, resulting in $%
\Delta _3(L)=1/3$. For energies just above the barrier height ($V<E<2V$, or $%
70<N(E)<110$) the situation is quite different. The rigidity seems closer to
the Wigner rigidity although no clear fit can be seen.

One can also consider an average of the rigidity over an ensemble of systems
with slightly different barrier heights. There are two main reasons for
considering an ensemble. First, since the region just above the barrier
contains only about forty levels which is relatively a small number of
levels, statistics over a larger number of levels are most desirable.
Another reason is that the averaging over different barrier heights may be
thought of as a time average over the energy levels of a system in which the
barrier height oscillates with a low frequency \cite{r8}. It is well known in
classical chaos that such a time dependent potential is necessary for the
development of chaos in one-dimensional systems. Therefore, one might expect
the ensemble averaged rigidity to approach more closely the Wigner form. The
rigidity after an ensemble average over 20 different barrier heights equally
spaced between $V=4.8$ and $V=5.2$ is indicated by the $\bigtriangleup $
symbols in Fig. \ref{fig.2}. It can be seen that the resulting rigidity is
closer to the Wigner rigidity. Since one may expect that some remnant of the
picket fence statistics will linger in the region just above the barrier
height, it makes sense to try a fit to the rigidity a function of the form $%
\Delta _3(L)=c_1(\ln (L)-.0687)/\pi ^2+c_2$. As can be seen in the figure,
for $c_1=0.58$ and $c_2=0.1$, a perfect fit is obtained.

In Fig. \ref{fig.3} the level spacing distributions for the ensemble
averaged case is presented. The high energies retain the picket fence
distribution while the low energies retain the double picket fence
distribution. The intermediate regime shows a distribution which is more
akin to the Wigner one. Nevertheless, this distribution is too messy for a
more precise statement, since the level spacing distribution is a less
precise tool than the spectral rigidity.

It is interesting that the Wigner type distribution is found for energies
just above the barrier height while the most complex behavior of the wave
function is observed for much lower energies \cite{r1}. We therefore
conjecture that the usual energy level statistics criteria correspond to a
signature of chaotic behavior in energy regions for which the classical
behavior of the system is similar to its quantum behavior. For energies deep
in the well, where the classical behavior is totally different than the
quantum one, the level statistics do not show a strong evident signature of
chaos from the point of view of the usual criteria. However, other, perhaps
more delicate tests should be investigated. For example, the double-picket
structure of the low-lying spectrum is only approximate. Our results \cite
{r1} in the study of the wave function with support in the neighborhood of $%
\sim \frac 1{10}$ of the barrier height showed very complex behavior.
Lewenkopf \cite{r7} has found that significant deviations occur for a $%
\delta $ - function barrier.

We have, furthermore, investigated the case in which the
barrier is shifted slightly off-center (the size of the regions
outside the barrier were chosen to be incommensurate so that
there exist no symmetry classes). We find that the results are
essentially identical, and therefore do not depend qualitatively
on the symmetry of the example that we have given.

One may also note that classical Hamiltonian chaos usually occurs in the
neighborhood of a separatrix, which in our case corresponds to energies in
the neighborhood of the barrier height. This is also in agreement with the
study of Pattanyak and Schieve \cite{r3} who have shown that, for a double
well potential, squeezed coherent states show the most complex behavior for
energies close to the separatrix. They argue \cite{r9} that the interplay
between the classically unstable orbits and the quantum tunneling effects is
the origin of this complex behavior. This is in agreement with our
observations on the behavior of the energy level statistics in that regime.

In conclusion, we have calculated the energy level statistics of a square
barrier embedded in an infinite well. For energies just above the barrier
level we find statistics which are closer to the Wigner statistics than in
other regions of the spectrum. When the spectral rigidity is averaged over
several different barrier heights the correspondence with the Wigner
prediction is very good. Thus a double well system exhibits signs of chaos,
for energies close to the separatrix.

R.B. would like to thank the Allon Foundation and the US--Israel Binational
Science Foundation for financial support.

\begin{figure}[tbp]
\caption{The cumulative number of levels $N(E)$ as function of energy. }
\label{fig.1}
\end{figure}

\begin{figure}[tbp]
\caption{The spectral rigidity $\Delta _3(L)$ as function of the energy
interval $L$ for different regions of the spectrum. The full curve
corresponds to the Wigner predictions $\Delta _3(L)=(\ln (L)-.0687)/\pi ^2$
, while the dotted curve corresponds to $\Delta _3(L)=0.58(\ln
(L)-.0687)/\pi ^2+0.1$. }
\label{fig.2}
\end{figure}

\begin{figure}[tbp]
\caption{ The level spacing distribution for different regions of the
spectrum. }
\label{fig.3}
\end{figure}

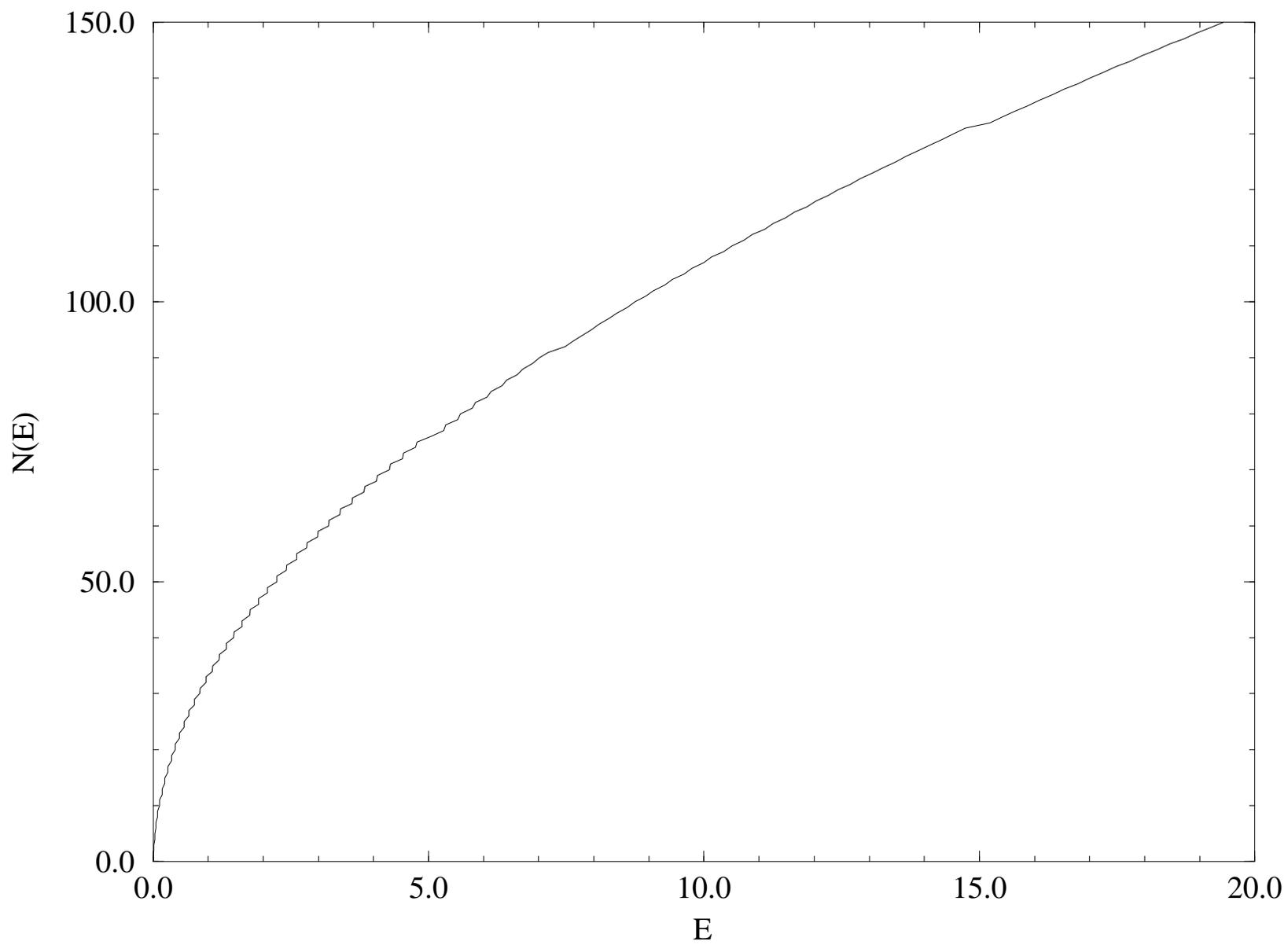
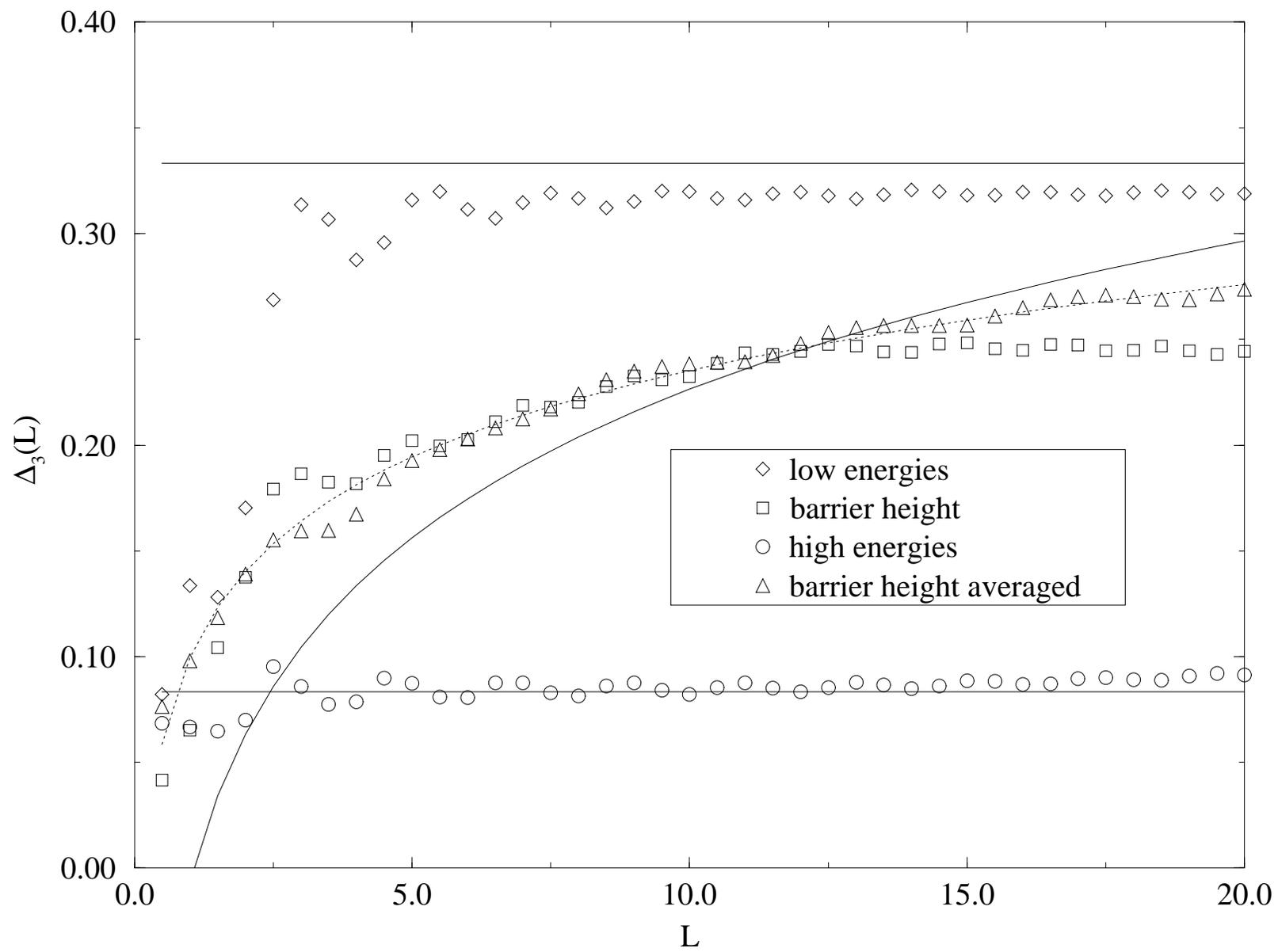
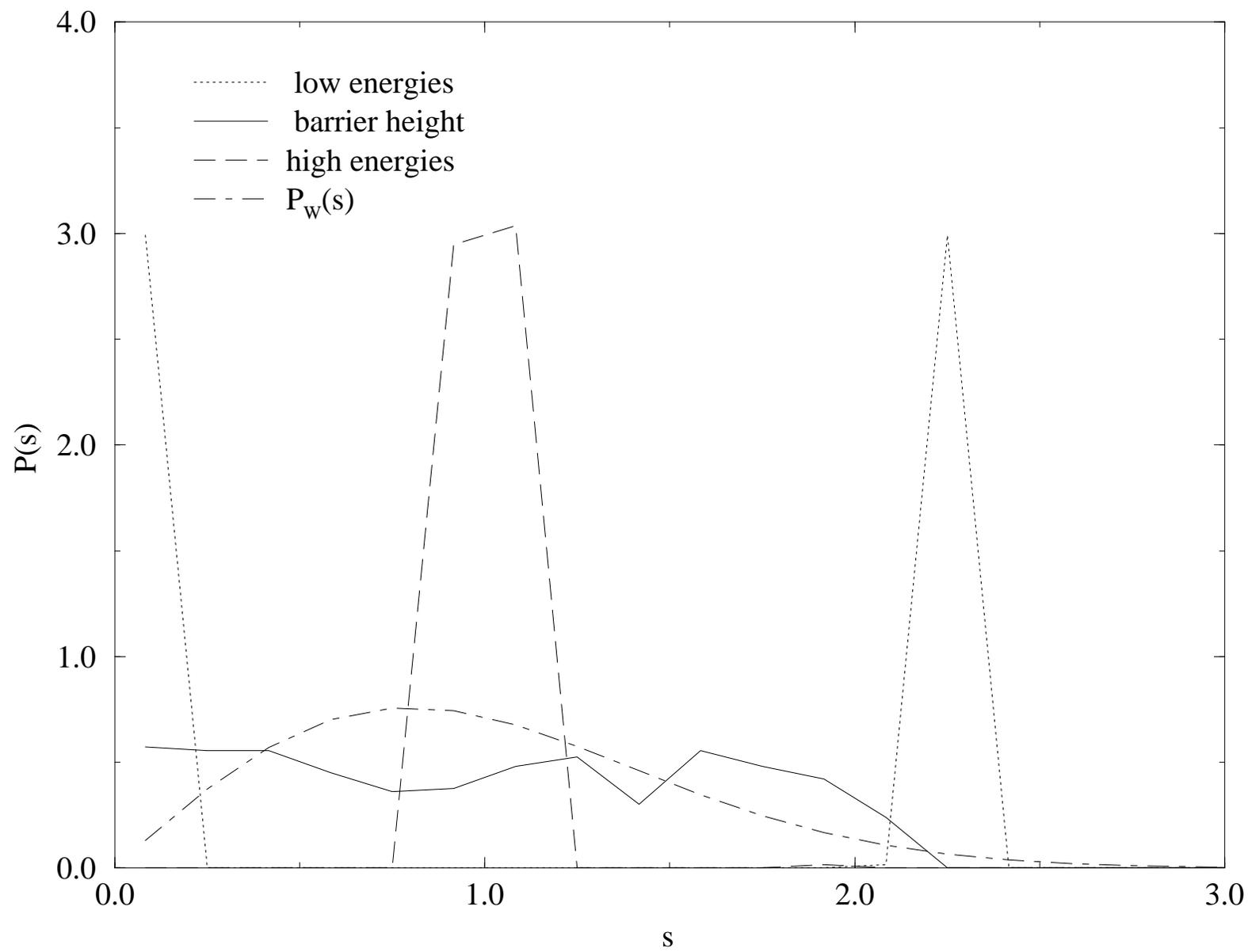
\end{document}